\documentstyle{article}
\begin{document}
\title{
Origin of GRB Afterglows in the Model of Galactic Neutron Stars}

\date{}
\author{Bisnovatyi-Kogan G.S.\thanks
{Space Research Institute, 84/32, Profsoyuznaya st.,
Moscow, Russia,\qquad
E-Mail: gkogan@mx.iki.rssi.ru}}
\maketitle

\begin{abstract}
The launch of the Beppo-Sax satellite gave a unique opportunity to investigate
gamma ray bursts (GRB) in different spectral regions. The large diversity of
the afterglow behavior creates additional problems for the cosmological
model with a fireball. Formation of the afterglow giving
the observed diversity of properties is suggested in the Galactic neutron
star model of GRBs. It is based on the transient accretion disc
formation around the neutron star with a low-mass brown companion
irradiated by the neutron star.
\end{abstract}
\section{Introduction}

A cosmological origin of GRBs leads to the conclusion of a huge energy output.
If the identification with the galaxy having redshift $z=3.42$ is true,
then in gamma radiation the energy release (without beaming) is
 $\sim 3 \times 10^{53}$ ergs (Kulkarni, 1998). Note, that the
solar rest mass is equal to  $1.8 \times 10^{54}$ ergs. Small timescale
indicates that GRBs are related to neutron
stars and (or) stellar mass black holes. The common event
from a collapse with a neutron star formation is a supernova explosion.
The total energy release $E_{tot}$ in a SN is equal to the
binding energy of a
neutron star $\sim 20\% Mc^2$. For the neutron star with a mass
$1.4\,M_{\odot}$ we get $E_{tot}\approx 5 \times 10^{53}$ ergs, which is
comparable to the above estimate for a GRB. Only a small
part $\sim 3 \times 10^{51}$ ergs is transformed into
kinetic energy of the explosion, and the energy radiated in all parts of the
electromagnetic spectrum is several tens times less. More then $99\%$
of the total energy output is emitted in the form of weakly interacting
neutrinos, and is dispersed in the Universe.
An artificially constructed
low-temperature structure around a neutron star or a black
hole may suffer from an instability. Stars with a neutron core
(Thorne-Zytkov model) are in most cases unstable to run-away neutrino
emission, leading to radiation by neutrinos of
more then $99\%$ of the accretion energy.
Magnetorotational explosion, used by Pazcynski (1998) to explain the
huge energy production in a cosmological GRB, had been suggested
for the supernova explosion by Bisnovatyi-Kogan (1971). Numerical 1-D
and 2-D calculations gave the efficiency of a
transformation of the rotational energy into the kinetic one at the
level of few percent (Ardelyan et al., 1997). The restrictions of the
"hypernova" model of Pazcynski (1998)  had been analyzed by
Blinnikov and Postnov (1998). The total
explosive energy output at the end of
the evolution of a close binary, consisting
of two neutron stars, suggested for a GRB model by Blinnikov at al.(1984)
cannot exceed the value of a (positive) binding energy less than
$10^{-3}M_{\odot}c^2= 1.6\times 10^{51}$ ergs (Saakyan and Vartanyan, 1964).
Only part of this energy may be radiated in the GRB region.
In the presence of serious energy problems inherent in the model of
the cosmological GRB, we try to explain the pioneering
results of the afterglow measurements by Beppo-Sax, as
well as previous hard gamma-ray afterglow observed by EGRET, in the
frame of the model of GRB origin in the old nearby neutron stars
inside the Galactic disk. The host galaxies with a high redshift are supposed
to be a chance coincidence with GRB. Isotropy on the sky and non 3/2
$\log N/\log S$ may be connected with selection effects (Bisnovatyi-Kogan,
1997), or a local non-uniformity (B.V.Komberg, priv. comm.).

\section{The model and neutron star statistics}

We consider a gamma-ray burst model based nuclear explosion
under the surface of a neutron star. There is a nonequilibrium layer
in the crust of the neutron star, consisting of very heavy and neutron
overabundant nuclei. After the starquake these nuclei are moved out, and
become unstable to fission after several beta-decays. This results
in an almost instant explosion and
the formation of a GRB (Bisnovatyi-Kogan et al.,
1975). The energy resource in the nonequilibrium layer is of the order
of $10^{47}$ ergs. If GRB originats at a distance about 100 pc with
the energy release $10^{36}-10^{40}$ ergs, then each neutron star may
give $10^7-10^{11}$ gamma ray bursts. Estimating the total number of neutron
stars in the Galaxy as (2-7)$\times 10^8$, and their number inside the sphere
with a radius 200 pc as (3-10)$\times 10^4$, we need a recurrence
time (100-300) years in old neutron stars to produce an observed GRB
every day (Bisnovatyi-Kogan, 1991).

\section{Hard X-ray afterglow}

The 1.5 hours afterglow observed by EGRET in GRB of 17 February 1994
in the energy band (40-18000) MeV has a simple explanation in the
local model. A nuclear explosion in the neutron star crust excites
eigen-oscillations in the neutron star with periods (0.1-10) ms. In presence
of such high frequency oscillations the old slowly rotating neutron star
with a relatively small magnetic field becomes a transient high
frequency radio pulsar. The radio pulsars are known as strongest sources in
the hard gamma ray band of EGRET, so the observed hard gamma ray afterglow
could be related to the activity of this transient high frequency radio pulsar
(Bisnovatyi-Kogan, 1995, 1997). To test this scheme it was suggested
to perform high time resolution radio observations of the Vela pulsar
shortly after the observed glitch, which happen almost every year.
This proposal (Bisnovatyi-Kogan and Tsarevsky, 1998) is now under
consideration.

\section{X-ray, optical and radio afterglows}

Beppo-Sax observations gave the X-ray flux for GRB afterglow on the level
$\sim 10^{-13}$ ergs/s. Taking 3 days for the duration of X-ray emission
we get $F_{\rm x}\sim 3\times 10^{-8}$ ergs/cm$^2$
for the X-ray fluence, which gives
$\sim 1/30$ of the main GRB with total fluency $\sim 10^{-6}$
ergs/cm$^2$. Explosion on the neutron star could lead to nonradial mass
ejection with a velocity between escape and Keplerian velocities. The ejected
matter falls back to the neutron star forming
an accretion disk due to its high
angular momentum. Taking the distance $\sim 100$ pc, corresponding to
the total GRB energy $E_{\gamma}\sim 10^{36}$ ergs, and $E_{\rm x} \sim
3\times 10^{34}$ ergs, we estimate the mass creating this X-ray flux
during accretion into a neutron star as $10^{-19}M_{\odot}$.
The spectrum of accretion with a low rate $\dot M \sim 3\times 10^{-17}
M_{\odot}$/yr$\approx 7\times 10^8$g/s, consist of two approximately equal
parts. The first is a relatively hard X-ray emission in the range 1-10 keV
from the boundary layer. The second is the radiation from the accretion disk
itself which emits a spectrum $F_{\gamma} \sim \omega^{1/3}$ with an
exponential cutoff starting from $\omega\approx 2\times 10^{15}$s$^{-1}$,
corresponding to a maximum temperature $2\times 10^4$ K, which color
corresponds to a $B$ star. Duration of the emission from the accretion
disk is not expected to last a long time, and it is not expected to
radiate much in the red and far red ranges, lasting several months
(Sokolov, 1998).

 The extended GRB afterglows with spectra corresponding to a cold star
may be explained easily if the neutron star has a low mass companion
with $M_d=(0.02-0.2) \,\,M_{\odot}$, as observed in most binary recycled
pulsars. Taking a companion with a mass 0.03$M_{\odot}$ and very low
temperature (degenerate brown dwarf) with a normal composition $(R_d\sim
5.3\times 10^9$ cm), or without hydrogen $(R_d\sim 2.2\times 10^9$ cm),
and binary separation $R_{12}\sim 10^{11}$ cm, we get a
time for binary merging
due to a gravitational radiation $\tau_g=\frac{5c^5 R_{12}^4}
{256 G^3 M_d M_n (M_d+M_n)}$=$2\times 10^{16} \frac{R_{11}}
{(M_d M_n (M_d+M_n))_{\odot}}$ sec. of the order of a cosmological time.
This companion absorbs $\sim(R_d/R_{12})^2$ of the total energy flux,
what is equal to $\sim 0.003$, $\sim 0.0005$ for normal composition, and
hydrogen-free dwarfs relatively. To obtain an afterglow with energy
comparable with the energy of the observed GRB we should imagine that
efficiency of GR production in the event does not exceed $\sim 1\%$, and
the main energy is radiated in the form of the kinetic energy, or
relativistic particles. It corresponds to the total energy output
$\sim 10^{38}$ ergs. The shock wave may heat the surface to high
temperatures, leading to an X-ray flash $\sim(R_{12}/c) \approx 3$ seconds
after the main GRB. For a longer GRB it means the corresponding change
of the GRB spectrum with a sharp rise in the soft part. Radiation flux
and ultrarelativistic particles penetrate deeper under the surface of
the star heating a rather thick layer. Taking the absorption cross-section
$\sim 10^{-26}$ cm$^2$, and the absorbed flux $ (1-3)\times 10^{35}$ ergs,
corresponding to $\sim 10^{16}$ ergs/cm$^2$, we obtain a
surface temperature
$\sim 10^5$ K immediately after absorption. It means that
relatively strong ultraviolet source appears, accompanied by a strong
mass loss feeding additionally the accretion disk around the neutron star.
After a short ($\sim 10$ seconds) phase of the mass loss and UV emission
the temperature drops, and a week cooling object appears with a spectrum
moving into the red and IR region, which is in general
accordance with the observed afterglows.

\end{document}